\documentclass[10pt,letterpaper]{article}
\usepackage{opex3}

\begin{document}

\title{Generation of squeezed light with monolithic optical parametric oscillator: Simultaneous achievement of phase matching and cavity resonance by temperature control
}

\author{Hidehiro Yonezawa, Koyo Nagashima, and Akira Furusawa}

\address{Department of Applied Physics, School of Engineering, The University of Tokyo, 7-3-1 Hongo, Bunkyo-ku, Tokyo 113-8656, Japan}

\email{yonezawa@ap.t.u-tokyo.ac.jp} 



\begin{abstract} 
We generate squeezed state of light at 860 nm with a monolithic optical parametric oscillator. 
The optical parametric oscillator consists of a periodically poled KTiOPO$_4$ crystal, both ends of which are spherically polished and mirror-coated. 
We achieve both phase matching and cavity resonance by controlling only the temperature of the crystal. 
We observe up to $-$8.0$\pm0.2$ dB of squeezing with the bandwidth of 142 MHz. 
Our technique makes it possible to drive many monolithic cavities simultaneously by a single laser. Hence our monolithic optical parametric oscillator is quite suitable to continuous-variable quantum information experiments where we need a large number of highly squeezed light beams.
\end{abstract}

\ocis{(270.6570) Squeezed states, (120.2920) Homodyning.} 


\section{Introduction}

Squeezed light is an important resource in various fields, e.g., precision measurement like gravitational wave detection \cite{Caves81,Goda08} and photonic quantum information processing with continuous-variables (CVs) \cite{Braunstein03,Cerf07}. 
In particular, a large number of squeezed light beams are often necessary for advanced applications in CV quantum information processing.
Generating and manipulating a number of squeezed light beams, however, is still a challenging task. 
Only recently several experiments with four or more squeezed light beams have been demonstrated. 
For examples, CV one-way quantum computation was recently demonstrated by using four squeezed light beams \cite{Ukai10}, and CV quantum error correction was realized with eight squeezed light beams \cite{Aoki09}. 
In order for such applications, it is essential to generate a large number of squeezed light beams simultaneously. Hence our aim is to develop a squeezer and technique by which we can generate a number of squeezed light beams simultaneously.

Squeezed light can be generated in various methods \cite{Slusher85,Wu86,Shelby86}.
In recent works, subthreshold optical parametric oscillator (OPO) is the most commonly-used squeezer \cite{Suzuki_Sq(2006),Takeno07,Goda_Sq08,Vahlbruch08,Masada_LN(2010),Mehmet10,Vahlbruch10}. So far the highest squeezing level was reported $-$11.5 dB by Mehmet \textit{et al.} \cite{Mehmet10}. They used monolithic OPO where two end surfaces of MgO:LiNbO$_3$ are spherically polished and mirror-coated. Thanks to the extremely low intra-cavity loss, which is an advantage of monolithic OPO, they observed such a highly squeezing at 1064 nm. 
Besides, they also demonstrated a broad spectrum of the squeezing over 170 MHz. Note that squeezing bandwidth of a conventional bow-tie OPO is limited to around 10 MHz \cite{Takeno07}. 
From these results, monolithic OPO would be the best choice of squeezers for CV quantum information experiments.  
There is, however, a serious problem when we apply the monolithic OPO to CV quantum information experiments. The monolithic OPO has a difficulty in controlling the cavity length. Mehmet \textit{et al.} did not lock the OPO cavity length \cite{Mehmet10} which is the same situation for previous monolithic OPO experiments \cite{Vahlbruch08,PKLAM_Sq(1999)}. 
Instead, they tune laser frequency to match a resonance condition of the OPO cavity. They cannot use several monolithic OPOs simultaneously which must be driven by the same laser. 
Therefore it is necessary to develop a technique for driving many monolithic OPOs simultaneously by a single laser. By developing such a technique, the monolithic OPO will be a promising squeezer for CV quantum information experiments.

In this paper, we demonstrate generation of squeezed lights with periodically poled KTiOPO$_4$ (PPKTP) monolithic OPOs. 
By controlling only the temperature of the crystal, we achieve both phase matching and cavity resonance of the monolithic OPO. 
We generate $-$8.0$\pm0.2$ dB of squeezing with a squeezing spectrum over 142 MHz. 
Our technique enables us to drive many monolithic OPOs simultaneously. Moreover our setup is quite simple and stable compared to the previous experiments. 
Hence our PPKTP monolithic OPO is easily applied to future quantum information experiments in which several broadband and high-level squeezed lights are required.

\section{PPKTP monolithic OPO}

In our experiments, we choose PPKTP crystal and use continuous wave laser at 860 nm according to Ref. \cite{Takeno07}.
Our PPKTP monolithic OPOs are provided by Raicol Crystals. The PPKTP crystals are 10 mm long with 1 mm $\times$ 1 mm cross-section. Both ends of the crystal are spherically polished (radius of curvature is 10 mm). Beam waist size inside the cavity is around 30 $\mu$m. One end of the crystal has a high reflection coating, and the other has a partial transmittance coating for 860 nm. We investigate three monolithic OPOs which have different partial transmittance coating, 11.8 \% (labeled as No.1), 8.2 \% (No.2) and 4.4 \% (No.3). For a second harmonic beam (430 nm), both ends have an anti-reflection coating. 
In other words, our OPOs are resonant only on the fundamental laser frequency. We do not use a doubly resonant OPO as in Ref. \cite{Mehmet10} because of the following reasons; (i) the temperature range of phase matching is narrower than that of a singly resonant OPO. (ii) PPKTP has strong absorption of blue light. It would be difficult to control the temperature of a doubly resonant OPO of PPKTP.

The PPKTP crystal is $a$-cut for both fundamental and second harmonic beams polarized along the $c$-axis. The poling period is 4.3 $\mu m$, and noncritical phase matching is achieved around 40 $^\circ$C. 
In order to fulfill both resonance and phase matching conditions simultaneously, we control only the temperature of the crystal. 
Phase matching is characterized by conversion efficiency $\eta$ from fundamental to second harmonic beam as \cite{Yariv_OE(1997)}, 
\begin{eqnarray}
\eta &\propto& \sin^2 (\Delta k l/2)/(\Delta k l  / 2 )^2, \\
\Delta k &=& \left ( \Delta n_Z^{(2\omega)} - \Delta n_Z^{(\omega)} \right ) 
           4 \pi / \lambda. 
\end{eqnarray}
Here $l$ is the crystal length, and $\lambda$ is the wavelength of the fundamental beam. $\Delta n_Z^{(2\omega)}$ and $\Delta n_Z^{(\omega)}$ are deviations of refractive indices from phase matching condition. These refractive indices are along $Z(c)$-axis for the second harmonic beam (2$\omega$) and the fundamental beam ($\omega$), respectively. Roughly speaking, the phase matching is achieved (that is, $\eta \geq 0.5$) under the condition of $\left | \Delta k l \right| \leq \pi$. 
Then the temperature range of phase matching $\Delta T_{\rm PM}$ can be calculated as, 
   \begin{equation}
      \Delta T_{\rm PM} \approx 
              \frac{\lambda}{2l} 
                     \left| 
                        \frac {dn_Z^{(2\omega)} }{dT} 
                      - \frac{dn_Z^{(\omega)} }{dT} 
                    \right|^{-1}. 
   \end{equation}
Here $dn_Z^{(2\omega)} /dT$ and $dn_Z^{(\omega)} / dT$ are given from Ref. \cite{Gurzadian_NOC(1999)} as $ 5.10 \times 10^{-5}$ ${\rm K^{-1}}$ and $ 3.57 \times 10^{-5}$ ${\rm K^{-1}}$, respectively. 
Note that if we choose a doubly resonant OPO, effective crystal length would be double and $\Delta T_{\rm PM}$ would be a half. 
The resonance condition is given by $2l n_Z^{(\omega)}=m\lambda$ where m is an integer. We can calculate a free spectrum range in terms of temperature $\Delta T_{\rm FSR}$ as, 
\begin{equation}
   \Delta T_{\rm FSR} = \frac{\lambda}{2l} \left| \frac{dn_Z^{(\omega)} }{dT} \right|^{-1}. 
\end{equation}
If $\Delta T_{\rm PM} > \Delta T_{\rm FSR} $, we can achieve both resonance and phase matching condition simultaneously by changing only the temperature. 

Figure. 1 shows theoretical calculations of temperature dependence of phase matching and resonance.
$\Delta T_{\rm PM}$ and $\Delta T_{\rm FSR}$ are calculated as 2.5 $^\circ$C and 1.2 $^\circ$C, respectively. We have at least two resonance peaks in the range of phase matching temperature. 
Although phase matching condition $\eta$ may not be perfectly optimized under the resonance condition, we can obtain {\it at least} 86 \% of $\eta$ compared to that in the perfect phase matching condition. 
Therefore we can achieve near-optimal condition by just controlling the temperature of the crystal.

     \begin{figure}[htbp]
       \centering\includegraphics[width=8cm]{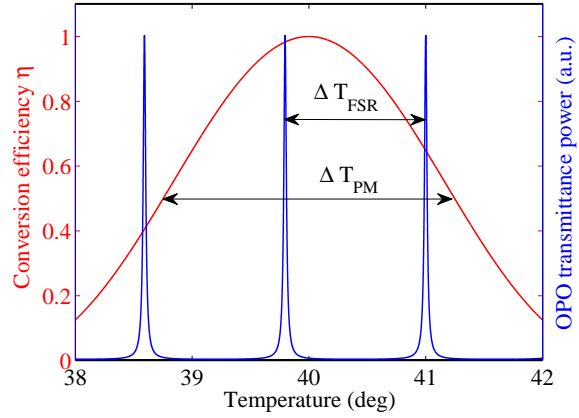}
        \caption{Theoretical calculation of phase matching and resonance condition. Blue line is OPO transmittance power (resonance condition). Red line is conversion efficiency $\eta$ from fundamental to second harmonic beam (phase matching condition). }
     \end{figure}

\section{Experimental setup}
Figure. 2 shows our experimental setup.
We use continuous wave Ti:sapphire laser (Coherent MBR-110) at 860 nm.
At first the beam is divided into three parts.
A fraction of the beam is used as a local oscillator (LO) beam for homodyne detection. In order to improve the fringe visibility between the LO beam and the squeezed beam, we use a mode-cleaning cavity (MCC). 
The most part of the beam is injected into a frequency doubler to generate second harmonic beam at 430 nm.
Here we use a bow-tie cavity with a LiNbO$_3$ crystal \cite{Masada_LN(2010)}, where 600 mW of 860 nm input is converted to around 250 mW of 430 nm second harmonic beam. This second harmonic beam is used to drive the OPO. The remaining 860 nm beam is utilized as a probe beam for several locking technique. 
We inject the probe beam into the monolithic OPO from the high-reflection-coated end. Here power of the probe beam is adjusted around 2 $\mu$W in the squeezed beam. Note that we need only one auxiliary beam for locking system while several auxiliary beams, which are often frequency-shifted, are usually used \cite{Suzuki_Sq(2006),Takeno07,Goda_Sq08,Masada_LN(2010),Vahlbruch10}. 
The generated squeezed light are interfered with the LO beam and measured by a homodyne detector. Then the homodyne detector output is recorded by a spectrum analyzer.
     \begin{figure}[htbp]
       \centering\includegraphics[width=10cm]{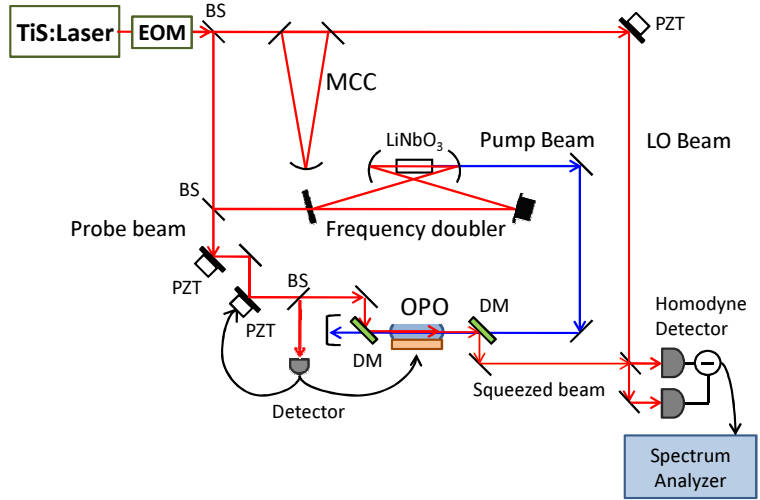}
        \caption{Experimental setup. Abbreviations are, EOM: electro-optic modulator, BSs: beam splitter, MCC: mode cleaning cavity, PZTs: piezoelectric transducer, and DMs: dichroic mirror (high reflection for 860 nm and high transmittance for 430 nm).}
     \end{figure}

We use standard Pound-Drever-Hall technique to lock the OPO cavity.
Probe beam is modulated at 36.7 MHz by an EOM in Fig. 2. The probe beam is injected into the OPO from the high-reflection-coated end. The reflected beam is measured by a detector as in Fig. 2. The output of the detector is demodulated and fed back through a PID controller to a peltier element attached to a copper crystal holder. 
Temperature is finely tuned because the full width at half maximum of the resonance is less than 0.03 $^\circ$C. Here temperature is controlled with a precision of 0.001 $^\circ$C around 40 $^\circ$C.

In order to measure an appropriate quadrature amplitude, we use two more locking techniques as Ref. \cite{Takeno07}. 
First we lock a relative phase between the pump beam and the probe beam. For this purpose, we use the same signal which is used to lock the OPO cavity. 
We modulate probe beam at 130 kHz in addition to 36.7 MHz by a PZT.
The output signal of the detector is split into two signals, one of which is already used for OPO locking. Another signal is demodulated with 130 kHz, and fed back to a PZT on a path of the probe beam. The error signal for this lock hardly suffers from fluctuation of the OPO cavity because we use much low frequency compared to the cavity bandwidth (82 MHz (No.1), 55 MHz (No.2) and 29 MHz (No.3) in terms of half width at half maximum). Note that the error signal for cavity locking, however, might greatly change according to parametric amplification. It is worth noting that we can cancel out this effect by adjusting phase of an electrical local oscillator used for the demodulation. 
After locking the relative phase between the pump beam and the probe beam, we lock relative phase between the probe beam and the LO beam by using homodyne output signal. Finally we can measure arbitrary quadrature amplitudes of squeezed light.

\section{Results and discussions}
First we measure squeezing with OPO No.1 (11.8 \% of an output coupler transmittance). 
Figure. 3 shows measurement results which are taken at center frequency of 2 MHz with resolution and video bandwidth of 30 kHz and 300 Hz, respectively. 
Here we use a homodyne detector which has high quantum efficiency (0.998) but slow frequency response (limited to around 6 MHz).  
The LO power is set as the largest where the linearity of shot noise power against LO power is verified. 
Here we use around 20 mW LO for measuring squeezing quadrature and 3 mW LO for measuring anti-squeezing quadrature. Note that we use lower LO power for measuring anti-squeezing quadrature because the large modulation signal is measured and it saturates the homodyne detector when we measure anti-squeezing quadrature. 
In the squeezing measurement, the shot noise level is 23 dB higher than the dark noise level at 2 MHz. 
The best squeezing is obtained at 130 mW pump power as shown in Fig. 3 (a).
We obtain $-$8.0$\pm0.2$ dB of squeezing and 16.0$\pm0.2$ dB of anti-squeezing.

   \begin{figure}[htbp]
    \centering\includegraphics[width=10cm]{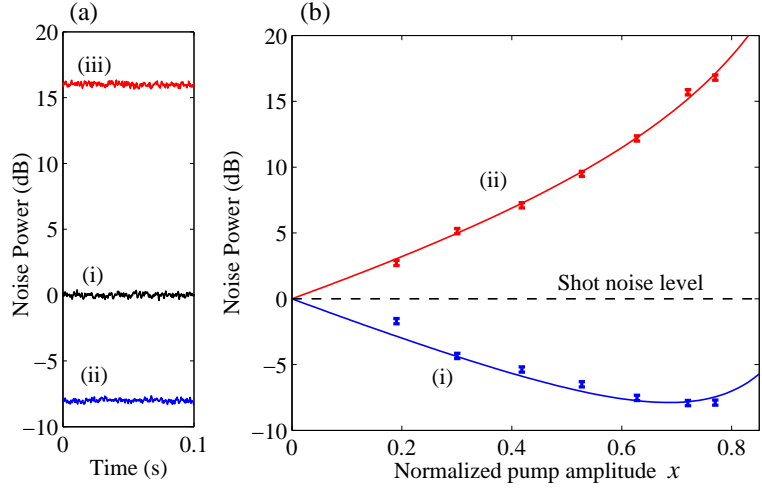}
    \caption{Measurement results of squeezed states. Here we use OPO No.1. All the data was measured at 2 MHz. (a) Measured at the fixed pump power (130 mW). Trace (i) is the normalized shot noise level. Trace (ii) and (iii) are the normalized squeezing and anti-squeezing level. (b) Pump power dependence of squeezing and anti-squeezing level. Trace(i) and (ii) correspond to the normalized squeezing and anti-squeezing level. Solid curves indicate the theoretical values. }
   \end{figure}

In order to analyze our squeezing, we measure squeezing with several pump powers. Figure. 3 (b) shows dependence of squeezing on a normalized amplitude $x$. Here we define $x=\sqrt{P/P_{th}}$ where $P$ and $P_{th}$ are pump power and the oscillation threshold power of the OPO, respectively. The threshold power $P_{th}$ is estimated as 283 mW from parametric amplification \cite{Takeno07}. 
In Fig. 3 (b), solid curves show theoretical values $R'_{\pm}$ which is given as \cite{Takeno07},  
  \begin{eqnarray}
     R'_{\pm} &\simeq& R_{\pm} \cos^2  \tilde \theta + 
                        R_{\mp} \sin^2  \tilde \theta , \\
     R_{\pm} &=& 1 \pm \kappa \frac{T}{T+L} 
                       \frac{4x}{(1 \mp x)^2 + (f/f_0)^2}.
     \label{sqtheory}
  \end{eqnarray}
Here the subscripts $+$ and $-$ stand for anti-squeezing and squeezing quadratures, respectively. $\kappa$ is overall propagation efficiency outside the OPO, $T$ is the output coupler transmittance, $L$ is the intra-cavity loss of the OPO, $f$ is the measurement frequency, $f_0$ is a frequency of half width at half maximum of the OPO, and $\tilde \theta$ is fluctuation of phase locking. Theoretical curves in Fig. 3(b) are fitted to the experimental data with the parameter $\tilde \theta$. $\tilde \theta$ is given as 2.0 degree which is consistent with 1.5 degree in Ref. \cite{Takeno07}. 
In our case a main factor in the limited squeezing level is an intra-cavity loss. The effect of the intra-cavity loss is characterized by an escape efficiency $T/(T+L)$. The escape efficiency is considered as an effective transmittance. The intra-cavity loss is measured 0.8 \% and then the escape efficiency is 0.937 which corresponds to 6.3 \% effective loss. 
The propagation coefficient $\kappa$ is given as 0.968 by taking into account homodyne visibility (0.986), propagation efficiency (0.998) and quantum efficiency of photo diode (0.998). Then the total loss outside the OPO is counted as 3.2 \%.
The intra-cavity loss may be caused by crystal defects or imperfect polishing. In a future work, it should be improved.

An advantage of monolithic OPO is capability of a broad spectrum of squeezing. We measure spectrum of squeezing up to 100 MHz as shown in Fig. 4.
In this measurement we use another homodyne detector which is tuned for broadband measurement but has a little lower quantum efficiency ($\sim0.99$). LO power is set as 10 mW for squeezing measurement and 3 mW for anti-squeezing measurement.
 In Fig. 4, squeezing and anti-squeezing level are normalized to the shot noise level. 
Here detector dark noise is subtracted. Note that the bandwidth of our homodyne detector is limited to around 40 MHz ($-$3dB bandwidth). In a high frequency region, the measurement is not accurate because of the low signal-to-noise ratio of the detector. Solid lines indicate theoretical curves which show good agreements with experimental results at least up to 40 MHz. Bandwidth of squeezing is defined as the frequency where the variance of squeezing $R_{-}$ increases up to $\frac{1}{2}\left( R_{-}^{min} +1 \right)$ from its lowest value $R_{-}^{min}$ \cite{Mehmet10}. In other words, we define squeezing bandwidth as a half width at half maximum of the Lorentzian function in the Eq. (\ref{sqtheory}), that is, $(1+x)f_0$. 
Estimated bandwidth of our squeezing is 142 MHz, which is more than ten times wider than that of a conventional bow-tie OPO \cite{Takeno07}.
   \begin{figure}[htbp]
    \centering\includegraphics[width=10cm]{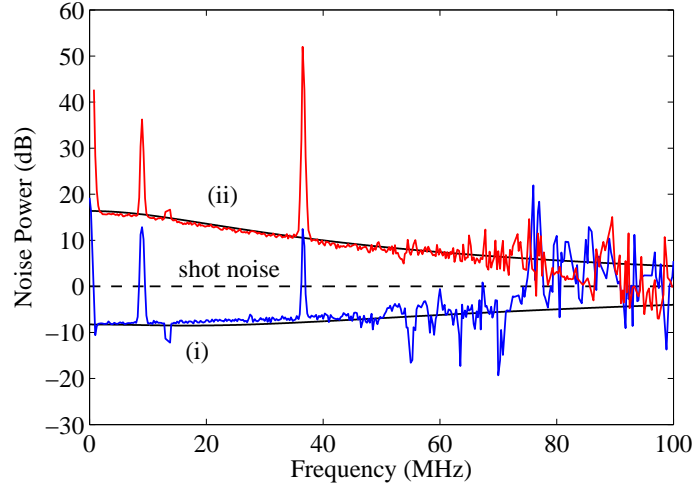}
    \caption{Broadband measurement of squeezed states. Trace (i) and (ii) correspond to the normalized squeezing and anti-squeezing level. Dark noise is subtracted. Solid black lines indicate the theoretical values. Spikes at 9.1 and 36.7 MHz are modulation signals for cavity locking.
}
   \end{figure}

Finally we mention about results of other crystals. We measure squeezing for OPO No.2 and No.3. The measured results are summarized in table \ref{tab1}. For comparison, we list the data of OPO No.1.
Particularly, oscillation threshold of OPO No.3 is very low (29 mW) although we still obtained $-$6.7$\pm0.2$ dB of squeezing with the bandwidth of 50 MHz. This low-threshold OPO is usable when we drive many OPOs simultaneously, because available pump power is limited in a real experimental situation.
Note that, in Ref. \cite{Mehmet10}, they used 600 mW of pump power to generate $-$11.5 dB of squeezing, and the threshold is estimated 1121 mW \cite{Masada_LN(2010)}.

	\begin{table}[h]
	\centering\caption{Measured data for three OPOs.} 
	\label{tab1}
        {\renewcommand\arraystretch{1.3}
	\begin{tabular}{ l|c|c|c } 
        \hline 
	Label			        &No. 1 & No.2  &No.3    \\ \hline
	Output coupler T & 11.8 \% & 8.2 \%  & 4.4 \%    \\ \hline 
	Threshold pump power & 283 mW & 92 mW  & 29 mW  
          \\ \hline 
	Squeezing at 2 MHz (dB) &$-$8.0$\pm0.2$ & $-$7.9$\pm0.2$ & $-$6.7$\pm0.2$ 
          \\ 
	(pump power) &(130 mW) & (60 mW) & (17 mW) 
          \\ \hline 
 	Anti-squeezing at 2 MHz (dB) &16.0$\pm0.2$ & 19.0$\pm0.2$ & 18.0$\pm0.2$
          \\
	(pump power) &(130 mW) & (60 mW) & (17 mW) 
          \\ \hline
 	Squeezing bandwidth & 142 MHz & 97 MHz  & 50 MHz    
          \\
	(pump power) &(130 mW) & (50 mW) & (15 mW) 
          \\ \hline

	\end{tabular}
	}
	\end{table}

\section{Conclusion}
We demonstrate generation of squeezed lights with PPKTP monolithic OPOs.
We achieve both phase matching and cavity resonance simultaneously by controlling the temperature of the crystal.
We generate $-$8.0$\pm0.2$ dB squeezing with the bandwidth of 142 MHz. Moreover one of our monolithic OPOs shows the low threshold power (29 mW) with yet high level squeezing ($-$6.7$\pm0.2$ dB) and broad spectrum (50 MHz). 
In the experiments, we control OPOs without changing laser frequency. Our technique enables us to drive many monolithic OPOs simultaneously by a single laser. Therefore our monolithic OPO will be especially useful for future CV quantum information experiments where we require several high level and broadband squeezed light beams. 

\section*{Acknowledgments}
Authors are grateful to Genta Masada for his experimental supports. H. Y. acknowledges M. Yukawa for helpful comments.
This work is partly supported by SCF, GIA, G-COE, PFN and FIRST commissioned by the
MEXT, RFOST, and SCOPE of the MIC.


\begin{thebibliography}{10}

 \bibitem{Caves81} 
  C. M. Caves, 
  ``Quantum-mechanical noise in an interferometer,'' 
         Phys. Rev. D {\bf 23}, 1693-1708 (1981).

 \bibitem{Goda08} 
  K. Goda, O. Miyakawa, E. E. Mikhailov, S. Saraf, R. Adhikari, K. Mckenzie,
  R. Ward, S. Vass, A. J. Weinstein, and N. Mavalvala, 
 ``A quantum-enhanced prototype gravitational-wave detector,''
  Nature Phys. {\bf 4}, 472-476 (2008).  

 \bibitem{Braunstein03} 
  S. L. Braunstein and A. K. Pati, 
 \textit{ Quantum Information with Continuous Variables} (Kluwer Academic Publishers, Dordrecht, 2003).

 \bibitem{Cerf07} 
  N. J. Cerf, G. Leuchs, and E. S. Polzik, 
 \textit{Quantum Information with Continuous Variables of Atoms and Light} 
(Imperial College Press, 2007).

 \bibitem{Ukai10} 
  R. Ukai, N. Iwata, Y. Shimokawa, S. C. Armstrong, A. Politi, J. Yoshikawa, 
  P. van Loock, and A. Furusawa, 
   ``Demonstration of unconditional one-way quantum computations 
     for continuous variables,''
  arXiv: 1001.4860 [quant-ph] (2010). 

 \bibitem{Aoki09} 
  T. Aoki, G. Takahashi, T. Kajiya, J. Yoshikawa, S. L. Braunstein,
  P. van Loock, and A. Furusawa,
   ``Quantum error correction beyond qubits,''
  Nature Phys. {\bf 5}, 541-546 (2009). 

 \bibitem{Slusher85} 
  R. E. Slusher, L. W. Hollberg, B. Yurke, J. C. Mertz, and J. F. Valley,
  ``Observation of Squeezed States Generated by Four-Wave Mixing in an 
    Optical Cavity,''
    Phys. Rev. Lett. {\bf 55}, 2409-2412 (1985).

 \bibitem{Wu86} 
  L. A. Wu, H. J. Kimble, J. L. Hall, and H. Wu,
   ``Generation of Squeezed States by Parametric Down Conversion,''
       Phys. Rev. Lett. {\bf 57}, 2520-2523 (1986).

 \bibitem{Shelby86} 
    R. M. Shelby, M. D. Levenson, S. H. Perlmutter, R. G. DeVoe, and D. F. Walls,
   ``Broad-Band Parametric Deamplification of Quantum Noise in an Optical Fiber,''
   Phys. Rev. Lett. {\bf 57}, 691-694 (1986).

 \bibitem{Suzuki_Sq(2006)} 
  S. Suzuki,  H. Yonezawa, F. Kannari, M. Sasaki, and A. Furusawa,
  ``7 dB quadrature squeezing at 860 nm with periodically poled KTiOPO$_4$,''
	Appl. Phys. Lett. {\bf 89}, 061116 (2006).

  \bibitem{Takeno07}
   Y. Takeno, M. Yukawa, H. Yonezawa, and A. Furusawa,
   ``Observation of -9~dB quadrature squeezing with improvement
    of phase stability in homodyne measurement,''
   Opt. Express {\bf 15}, 4321-4327 (2007).

  \bibitem{Goda_Sq08}
   K. Goda, E. E. Mikhailov, O. Miyakawa, S. Saraf, S. Vass, A. Weinstein, 
   and N. Mavalvala,
   ``Generation of a stable low-frequency squeezed vacuum field with 
     periodically poled KTiOPO$_4$ at 1064 nm,''
      Opt. Lett. {\bf 33}, 92-94 (2008).

  \bibitem{Vahlbruch08} 
   H. Vahlbruch, M. Mehmet, S. Chelkowski, B. Hage, A. Franzen, 
   N. Lastzka, S. Go\ss ler, K. Danzmann, and R. Schnabel,
   ``Observation of Squeezed Light with 10-dB Quantum-Noise Reduction,''
   Phys. Rev. Lett. {\bf 100}, 033602 (2008).

   \bibitem{Masada_LN(2010)} 
      G. Masada,  T. Suzudo, Y. Satoh, H. Ishizuki, T. Taira, and A. Furusawa,
    ``Efficient generation of highly squeezed light with periodically poled 
       MgO:LiNbO$_3$,''
      Opt. Express {\bf 18}, 13114-13121(2010).

 \bibitem{Mehmet10} 
    M. Mehmet, H. Vahlbruch, N. Lastzka, K. Danzmann, and R. Schnabel,
    ``Observation of squeezed states with strong photon-number oscillations,''
    Phys Rev A {\bf 81}, 013814 (2010).  

  \bibitem{Vahlbruch10} 
    H. Vahlbruch, A. Khalaidovski, N. Lastzka, C. Gr\"{a}f, 
    K. Danzmann, and R. Schnabel,
    ``The GEO600 squeezed light source,''
    Class. Quantum Grav. {\bf 27}, 084027 (2010). 

  \bibitem{PKLAM_Sq(1999)} 
   P. K. Lam, T. C. Ralph, B. C. Buchler, D. E. McClelland, H-A. Bachor, 
   and J. Gao,
   ``Optimization and transfer of vacuum squeezing from an optical 
     parametric oscillator,''
   J. Opt. B: Quantum Semiclass. Opt. {\bf 1}, 469-474 (1999).

  \bibitem{Yariv_OE(1997)}
  A. Yariv,
  \textit{Optical Electronics in Modern Communications 5th ed.}
  (Oxford University Press, Oxford New York, 1997).


  \bibitem{Gurzadian_NOC(1999)}
   G. G. Gurzadian, V. G. Dmitriev, and D. N. Nikogosian,
   \textit{Handbook of Nonlinear Optical Crystals}
   (Springer, 1999).

\end{thebibliography}
\end{document}